# Reconfigurable ultrafast perovskite polariton logic gates via nonlinear dynamics

Yuyang Zhang[1,#], Zhuoya Zhu[2,3,#], Xin Zeng[2], Shuai Zhang[2], Xinyi Deng[1], Tian Lan[1], Changhai Zhu[1], Kwok Kwan Tang[1], Qinglin Jia[1], Yuexing Xia[2,3], Yiyang Gong[2,3], Wenna Du[2,3], Feng Li[4], Rui Su[5], Xuekai Ma[6], Xinfeng Liu[2,3,*] and Qing Zhang[1,*]

[1]School of Materials Science and Engineering, Peking University, Beijing 100871, China

[2]CAS Key Laboratory of Standardization and Measurement for Nanotechnology, CAS Center for Excellence in Nanoscience, National Center for Nanoscience and Technology, Beijing 100190, China

[3]University of Chinese Academy of Sciences, Beijing 100049, China

[4]Key Laboratory for Physical Electronics and Devices of the Ministry of Education & Shaanxi Key Lab of Information Photonic Technique, School of Electronic Science and Engineering, Faculty of Electronic and Information Engineering, Xi'an Jiaotong University, Xi'an 710049, China

[5]Division of Physics and Applied Physics, School of Physical and Mathematical Sciences, Nanyang Technological University, Singapore 637371, Singapore

[6]Department of Physics and Center for Optoelectronics and Photonics Paderborn (CeOPP), Paderborn University, Warburger Strasse 100, Paderborn 33098, Germany

[*]Email address: q_zhang@pku.edu.cn; liuxf@nanoctr.cn;

[#]These authors contributed equally to this work.

**Abstract**

Exciton-polaritons provide a great platform for developing ultrafast all-optical logic gates for quantum and optical chips. However, progress toward practical polariton logic remains limited due to incomplete logical functionality on a single device. Herein, we present a single-device perovskite polariton platform enabling reconfigurable, ultrafast logic gates with functional completeness. The device consists of an optically trapped perovskite microwire, generating well-controlled non-equilibrium polariton condensation states for multiple logic operation channels. By tailoring the power of signal and gate beams, the same device is programmed to execute three basic Boolean functions ('AND', 'OR', and 'NOT') and a high-order 'XOR' function with a high on/off ratio of ~21 dB, and a fast response time ~6.7 ps. The reconfigurability arises from the selective activation of different nonlinear responses of polariton condensates, including amplification, seeding state transitions, and nonlinear interaction. These results provide valuable insights for advancing exciton-polariton logic gates.

Exciton-polaritons (EPs) are hybrid bosonic quasiparticles emerging from strong light-matter coupling between excitons and photons[1–5]. They inherit a small effective mass from photons and exhibit strong nonlinear interactions derived from excitons. These unique characteristics promote the non-equilibrium polariton condensates at ambient temperatures[6–11], which can be controlled on an ultrafast timescale[12–15], thereby providing a great platform for all-optical, high-speed binary logic operations[16–23], including ultrafast transistors, switches, modulators and logic gates[14,24–28]. However, the development of practical device-level EP logic remains challenging. Most reported schemes rely predominantly on ground-state condensation, which limits the accessible polariton channels and constrains the implementation of multifunctional logic within single device[14,18,21]. In addition, cascaded polaritonic logic typically relies on complex configurations, making compact and scalable integration difficult[16,20,22]. These challenges motivate the development of reconfigurable EP logic platforms capable of supporting logic-complete sets for ultrafast optical logic circuits.

Benefiting from their large exciton binding energy and high quantum emission yield, metal halide perovskites have emerged as promising candidates for room-temperature polariton platforms, enabling diverse phenomena including superfluidity, continuous-wave condensation, topological lasing, quantum simulators, and optical spin Hall effect[29–43]. Furthermore, their strong nonlinear interactions, combined with the ease of fabricating large-scale planar structures, provide an attractive basis for scalable ultrafast EP logic gates and circuits[34–37]. For example, one-dimensional perovskites, such as microwire and nanowire, can support efficient polariton transport for cascaded operations[38–40]. Together with the tailored pumping and potential-engineering schemes enabled by microfabrication, support well-controlled non-equilibrium and large-wavevector polariton condensation states for realizing complete logic functions in single device[41–43]. These features make halide perovskites attractive for addressing the constraints of existing EP logic platforms.

In this work, we demonstrate an optically trapped perovskite microwire cavity enabling single-device logic-complete polaritonic gates operating on a picosecond timescale. Four types of Boolean function including 'AND', 'OR', 'NOT' and 'XOR' have been realized with high on/off-ratio and cascadable output. The reconfigurable logic functionality stems from the threshold-related nonlinear dynamics of trapped EP condensates with discrete energy states. By tuning the power of signal beam (meantime serving an optical trap) and gate beam, the established amplification, seeding, and repulsive interaction of polaritons give rise to 'AND', 'NOT', and 'OR' / 'XOR' functions, respectively. These results provide important insights into the development of polaritonic gates.

**Fig. 1a** shows the schematic of the optically-trapped perovskite polariton logic gate. The cavity employs high-quality single-crystal perovskite microwires embedded between distributed Bragg reflectors (DBRs, Fig. S1-S3). Room-temperature EP condensation is achieved under non-resonant femtosecond excitation (Fig. S4-S6). The one-dimensional confinement of the microwire generates the trapped EPs with discrete and narrow-width energy levels as multiple logic operation channels. A pulsed ring-shaped signal beam, with the diameter smaller than the microwire length, is used to create an optical trap and modifies the predominant EP states (**Fig. 1b, 1c**, S7-S10). The condensates preferentially occur on the states matching the standing-wave modes confined by optical trap, which are quantized as $n$ = 1-5. A time-delayed spot-shaped beam is introduced as gate; as the both beams are applied, the configuration is defined as the '11' input state.

By tailoring the intensities and the time delay of signal/gate beams, we can control the nonlinear response of the trapped polaritons and thus reconfigure the logic functions, as shown in **Fig. 1d**. For the 'AND' logic function, both beams are set below the condensation threshold ($P_s$, $P_g < P_{th}$). Neither beam (input: '10' or '01') alone can induce condensation (output: '0'), whereas dual-beam excitation drives condensation (output: '1'). For the 'NOT' logic function, the signal beam intensity is set above threshold ($P_s > P_{th}$) to generate condensation at high-$n$ states, whereas the gate beam intensity remains below threshold ($P_g < P_{th}$). Acting as a seed, the weak gate beam drives the condensate towards low-$n$ states and switches off the original output from '1' to '0'. For the 'OR' and 'XOR' logic functions, both beams are set above threshold ($P_s$, $P_g > P_{th}$), so that condensation occurs under any active input state, satisfying the requirement for 'OR' logic. Meanwhile, the simultaneous high-power excitation generates a high EP density inside the potential well, strengthens the repulsive interactions, and induces a blueshift of the polariton states. By selecting an appropriate logic energy window, this interaction-induced blueshift shifts the condensate out of the predefined detection range under the '11' input state, thereby producing a distinct 'XOR' output. Furthermore, the condensates flow along the long-axis direction of microwire and enable the cascadability of output results (Fig. S11).

**Fig. 2a** demonstrates the time-resolved 'AND' logic operation, in which both the signal beam and gate beam are set below threshold ($P_s = 0.6\ P_{th}$ and $P_g = 0.6\ P_{th}$) with a controllable delay $\Delta t$ between the two excitation pulses. A positive time delay is defined when the signal beam arrives earlier than the gate beam. To clearly reveal the amplification dynamics underlying the 'AND' logic operation, we selected a small-spot signal beam (2 μm) to preferentially induce condensation near the $n = 1$ state. In this case, the single-state occupation allows a clear observation of amplification process, since either beam is insufficient to induce condensation and yields weak emission, corresponding to the logic output '0' (**Fig. 2b**). However, the signal beam establishes an exciton reservoir, thereby enabling stimulated amplification when the second beam arrives. This results in EP condensation at $n = 1$ state and produces an output transition from '0' to '1' (**Fig. 2c**). Furthermore, the delay-time-dependent output intensity shows a peak centered at $\Delta t = 0$ and a full width at half maximum of approximately 36.7 ps (**Fig. 2d**). This temporal profile reflects the transient behavior of the exciton reservoir, where the gain is maximized under synchronous excitation, enabling efficient stimulated scattering and condensation. As the time delay increases, the reservoir dissipates before the arrival of the second one, resulting in a gradual reduction of the amplification gain. As shown in **Fig. 2e**, a stronger gate beam raises exciton density and enhances stimulated amplification, resulting in a substantial enhancement of the logic output extinction ratio as a function of gate power, with a maximum on/off ratio of 21 dB.

For the 'NOT' logic operation (**Fig. 3a**), a high-power signal beam ($P_s = 1.4\ P_{th}$) was used to establish the high-$n$ condensate ($n = 5$), while a weak gate beam ($P_g = 0.4\ P_{th}$) was introduced to control the output. As the time delay decreases from 13.3 ps to -13.3 ps (**Fig. 3b**), the condensate states transit from $n = 5$ towards $n = 3$, reflecting that the gate beam induced seeding effect drives relaxation towards low-$n$ states. Accordingly, a clear 'NOT' switching from '1' to '0' is obtained at negative time delay, as evidenced by the suppression of the $n = 5$ state emission (**Fig. 3c**). **Fig. 3d** presents a maximum on/off ratio of 14 dB and a rapid decrease within 6.7 ps. Notably, an asymmetric response is resolved in time scale, which may provide an opportunity for directional temporal gating. This asymmetry reflects the time-sequence dependence of the seeding effect, where the earlier arrival of the gate beam promotes population of the $n = 1$ state and enhances the switching-off of the high-$n$ condensate. Meanwhile, the extinction ratio exhibits a non-monotonic dependence on gate power (**Fig. 3e**). This originates from the competition between enhanced stimulated relaxation and nonlinear blueshift: at a low gate power, increasing density promotes polariton relaxation toward low-$n$ states, improving the extinction ratio; while, further increasing the gate power induces a significant blueshift (~1 meV) of the energy levels, shifting the condensate out of the predefined energy window and thereby reducing the on/off ratio.

Furthermore, the realization of 'OR' and 'XOR' gates requires that the pump powers of both beams exceed the threshold ($P_s = 1.8\ P_{th}$ and $P_g = 1.1\ P_{th}$), ensuring that condensation occurs under any input condition to satisfy the 'OR' logic requirement (**Fig. 4a**). Simultaneously, high-power excitation activates strong nonlinear blueshift, enabling a well-defined 'XOR' functionality. As shown in **Fig. 4b and 4c**, under the '10' and '01' input states, both above-threshold excitations produce condensates. Under the '11' input state, the nonlinear blueshift lifts the energy of the $n = 1$ state. The system exhibits the strongest nonlinear interaction near zero delay, leading to a maximum blueshift of 1.5 meV and a highest on/off ratio of ~20 dB (**Fig. 4d**). An asymmetric evolution of blueshift is resolved with varying the time delay. The gate beam preferentially drives condensation into the $n = 1$ state, whereas the signal beam populates both the $n = 1$ and high-$n$ states. The temporal order of the two beams therefore determines the polariton density and blueshift of the $n = 1$ state. At negative time delay, where the gate beam arrives first, the $n = 1$ state maintains a high polariton density, sustaining a large blueshift even at $\Delta t = -40$ ps. At positive time delay, the population redistributes towards high-$n$ states. The reduced occupation of the $n = 1$ state weakens the nonlinear blueshift, which decays within a fast response time of 6.7 ps. Consistently, **Fig. 4e** shows the power-dependent evolution. The dependence of both the blueshift and the detection-window emission intensity on the gate beam power is consistent with the mean-field polariton-polariton interaction relationship predicted by the Gross-Pitaevskii model[44], confirming that the 'XOR' functionality originates from nonlinear interaction effects rather than the potential reshaping.

Next, we demonstrate reconfigurable logic operations of 'AND', 'NOT', 'OR', and 'XOR' within a single device. The implementation of the 'AND' logic function is demonstrated in **Fig.**

**5a**. The single beams are set below the condensation threshold ($P_s = 0.7\ P_{th}$, $P_g = 0.7\ P_{th}$), yielding only negligible background emission defined as logic output '0'. The cooperative excitation of both beams activates stimulated amplification, leading to the formation of polariton condensates at $n = 3$ state, serving as the logic '1' output. **Fig. 5b** shows the implementation of 'NOT' function. We set the signal beam intensity to $1.5\ P_{th}$, generating a condensate state localized at $n = 3$, 4, and 5. A gate beam is then introduced as a seed ($P_g = 0.7\ P_{th}$), which suppresses these states and realises the 'NOT' logic operation. **Fig. 5c,d** demonstrate the 'OR' and 'XOR' logic functions under the same excitation configuration ($P_s = 2.2\ P_{th}$, $P_g = 1.5\ P_{th}$). The full emission energy range gives the 'OR' response. The energy in the vicinity of the $n = 1$ state under single-beam excitation serves as the 'XOR' detection window. However, the blueshift induced by cooperative excitation (input: '11') shifts the $n = 1$ state out of the original detection window, serving as the logic '0' output and thus satisfying the 'XOR' response. The corresponding normalized emission intensities from 50 measurements further confirm the reproducibility of the four logic operations **(Fig. 5e)**.

**Conclusion**

In summary, we demonstrate reconfigurable, ultrafast logic gates with full functional completeness operating at room temperature in a perovskite microcavity. A single logic gate device implements four Boolean functions 'AND', 'OR', 'NOT' and 'XOR' with high on/off ratio and picosecond response time. The reconfigurable functions are accomplished by the nonlinear dynamic control of trapped polaritons. Moreover, cascadability is supported by the delivery of output logic signals along the microwire. These findings provide a pathway toward room-temperature polaritonic logic circuits, unlocking new possibilities for integrated devices.

**Methods**

*Materials preparations*: CsPbBr$_3$ microwires were prepared by a chemical vapor deposition (CVD) process. 10 mg CsBr and 20mg PbBr$_2$ powder (99.999%, Sigma-Aldrich) were mixed, loaded into a quartz boat, and placed in a tube furnace (Lindberg/Blue M TF55035C-1). Three silicon wafers (100 plane, 1 × 1 cm$^2$) were positioned 25 cm downstream from the quartz boat in the tube furnace. After evacuating the chamber for 2 hours, high-purity nitrogen (99.99%) was flowed at 30 sccm to stabilize the pressure at 400 Torr. The furnace temperature was ramped to 575 °C over 20 min and held for 10 min, yielding CsPbBr$_3$ nanowires on the Si substrates.

*Microcavity fabrications*: The microcavity was constructed by embedding the perovskite into two dielectric distributed Bragg reflector (DBR) fabricated by electron-beam evaporation. For the bottom DBR, 19.5 pairs of TiO$_2$ (54 nm)/SiO$_2$ (88 nm) layers were deposited on Si wafers, followed by an additional SiO$_2$ layer for cavity-length adjustment. The CsPbBr$_3$ microwires were then mechanically transferred from the Si substrate onto the bottom DBR by gentle tapping, thereby releasing the microwires onto the DBR surface. After transfer, a SiO$_2$ spacer layer was deposited as a protective and cavity-spacing layer. The cavity was finally completed by depositing the top DBR, consisting of 9.5 pairs of SiO$_2$/TiO$_2$ layers with the same thicknesses as those of the bottom DBR.

*Morphology Characterizations*: The thickness and surface roughness of the perovskite microwires were characterized by atomic force microscopy (AFM, Veeco Dimension 3100, tapping mode). Scanning electron microscopy (SEM) images were obtained using a Hitachi SU8220 system. Optical images were captured with an Olympus FV3000 microscope system.

*Optical Spectroscopy Characterizations*: Photoluminescence spectra were collected by focusing a 405 nm continuous-wave laser onto the sample surface through a ×100 objective (Olympus, NA = 0.9). The emitted signals were analysed using a liquid-nitrogen-cooled spectrometer (SP2500i, Princeton Instruments). Angle-resolved photoluminescence spectra were collected using a home-built Fourier-space imaging system. The schematic of the experimental setup is provided in Supplementary Figure S8. The 1 kHz pulsed lasers were focused onto the sample through a ×50 objective (Olympus, NA = 0.8), and the reflected signal was collected by a 4f Fourier-imaging system. After spatial purification using an adjustable aperture, the signal was directed to the same spectrometer for spectral analysis. The dual-pulse excitation scheme was implemented using a beam splitter and a delay line. The 400 nm laser was divided into two beams. The ring-shaped signal beam was generated by combining an axicon (AX2505, Thorlabs) with a Galilean telescope (BE02-UVB, Thorlabs). The spot-shaped gate beam was produced by focusing the laser through an f = 150 mm len together with a spatial aperture for beam selection. The temporal delay between the two beams was tuned using a delay line with a maximum travel range of 15 mm. For time-resolved photoluminescence, a 1030 nm pulsed laser operating at 10 kHz was frequency-doubled to 515 nm and used as the excitation source, while the emission dynamics were recorded using a time-correlated single-photon-counting spectrometer.

**Supplementary Information**

Supplementary Information is provided free of charge and contains additional morphological characterizations, spectral images, and schematics that support the discussions in the main text.

**Author Information**

Corresponding Author

*Email: q_zhang@pku.edu.cn; liuxf@nanoctr.cn;

**Notes**

The authors declare no competing financial interest.

**Acknowledgments**

The authors acknowledge funding support from the National Key R&D Program of China (2024YFA1208203, 2023YFA1507002), the National Natural Science Foundation of China (U23A2076, 22325301, 22073022 and 22173025), the Open Research Fund of State Key Laboratory of Quantum Functional Materials (QFM2025KF001), Beijing Natural Science Foundation (1262028) and the Strategic Priority Research Program of Chinese Academy of Sciences (XDB0770000).

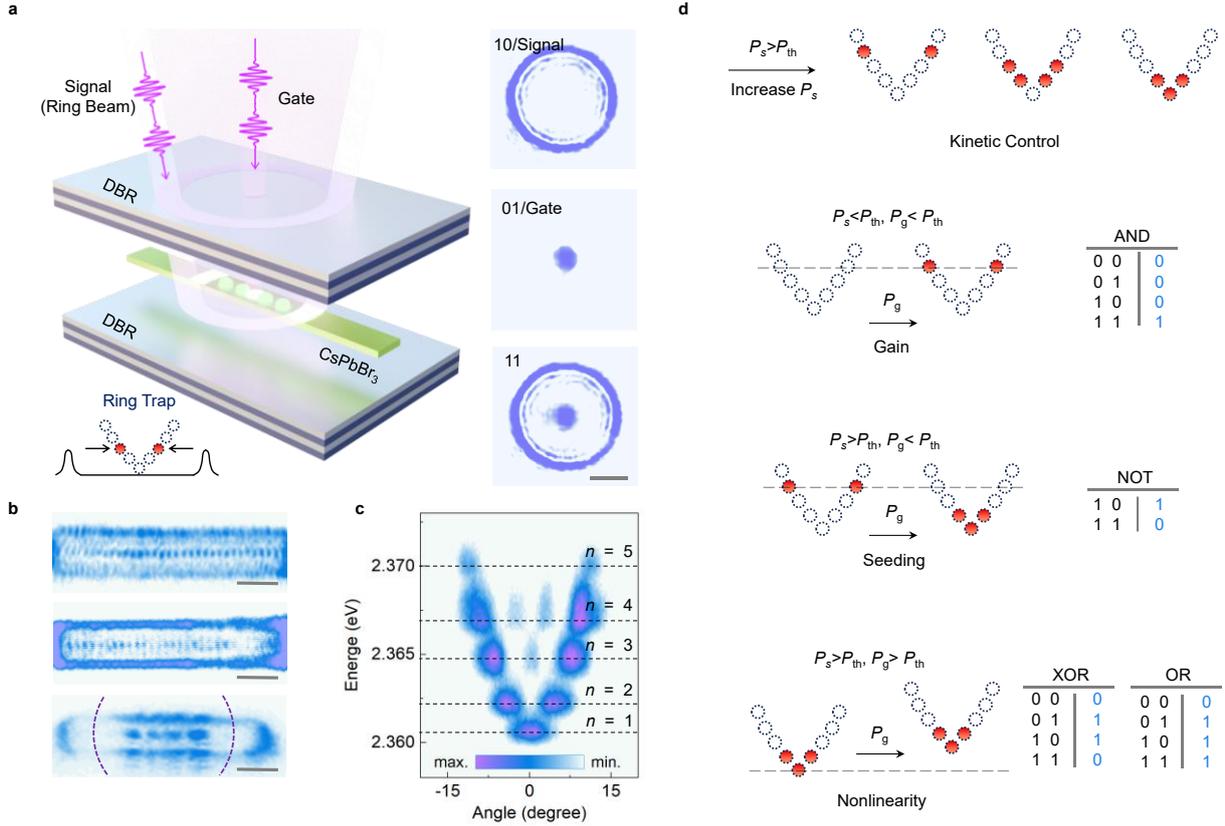

**Figure 1 Formation of one-dimensional exciton-polariton condensation and realization of reconfigurable logic gates in the CsPbBr$_3$ microwire cavity.** **(a)** Schematic illustration of one-dimensional EP condensation. Under non-resonant excitation by a ring-shaped signal beam (purple ring), repulsive interactions lead to exciton accumulation at the excitation region, forming a ring-shaped potential barrier. The counterpropagating EP flows confined within the inner sides of the potential barriers merge to form condensates (real space: green spheres; $k$-space: orange solid circles). A gate beam is subsequently applied to kinetically control the condensation. Right panel: Real-space optical images corresponding to the initial excitation configurations of the '10', '01', and '11' stages. Scale bar: 3 μm. **(b)** Real-space optical image of one-dimensional EP condensation. Large-radius ring excitation covering the entire microwire (top). Michelson interferometric image revealing the spatial coherence of the condensate (middle). Ring-shaped signal beam excitation (bottom). The black dashed line indicates the position of the excitation ring. Scale bar: 3 μm. **(c)** Angle-resolved photoluminescence (ARPL) spectrum of condensed one-dimensional EPs at $P_s = 2.2\,P_{th}$. The gray dashed lines denote different discrete polariton states. **(d)** Illustration of the mechanisms enabling reconfigurable Boolean logic operations, including kinetic control, stimulated amplification, seeding, and nonlinear interaction within EP system. Gray dashed circles represent discrete energy levels, and orange solid circles denote condensates formed on the corresponding levels. The light-colored solid lines indicate the energy-collection regions for the different logic gates.

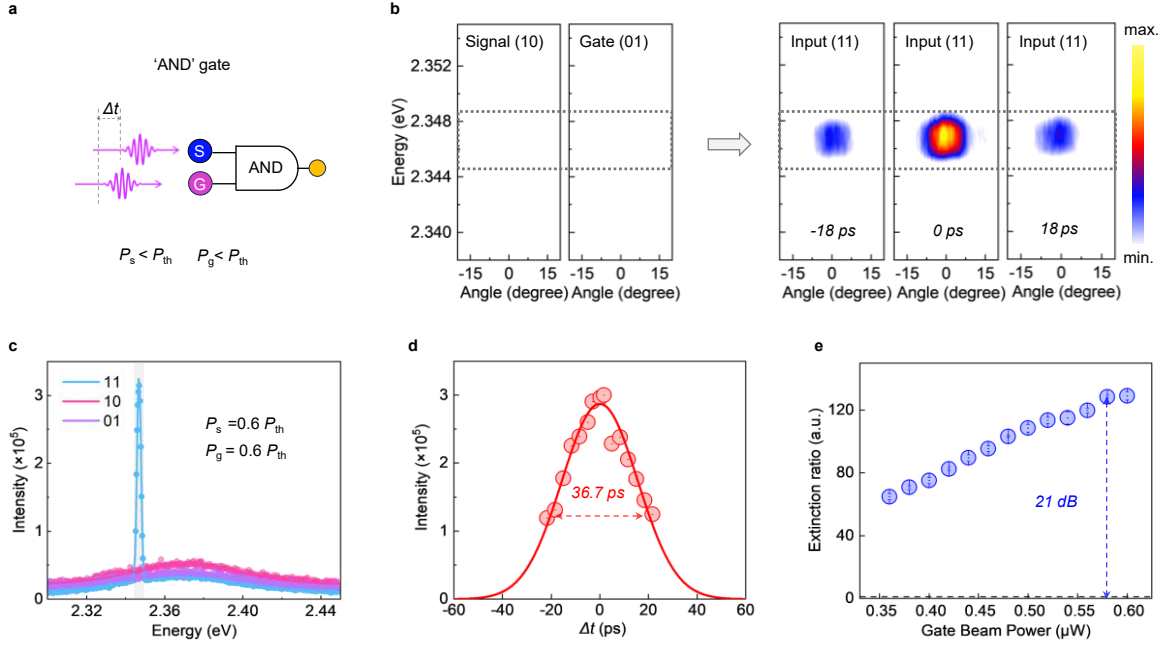

**Figure 2 Ultrafast exciton-polariton 'AND' logic operation. (a)** Schematic illustration of the 'AND' gate mechanism. Both the signal and gate beams are set below the condensation threshold. **(b)** ARPL spectra at different time delays between the signal and gate beams, revealing the temporal evolution of the stimulated amplification process underlying the 'AND' logic operation. The left panels correspond to excitation by the signal beam only ($P_s$ = 0.6 $P_{th}$) and the gate beam only ($P_g$ = 0.6 $P_{th}$), respectively, while the right panels show the condensate distributions at time delays of -18 ps, 0 ps, and 18 ps. The black dashed box marks the collection region corresponding to the 'AND' gate. **(c)** Output of the 'AND' logic gate under different input states. The gray shaded region corresponds to the energy collection window in (b). **(d)** Peak emission intensity as a function of time delay, showing maximum amplification near zero delay with a full width at half maximum of ~36.7 ps. **(e)** Extinction ratio as a function of gate beam power, exhibiting a linear increase below the threshold, with a maximum extinction ratio of 21 dB.

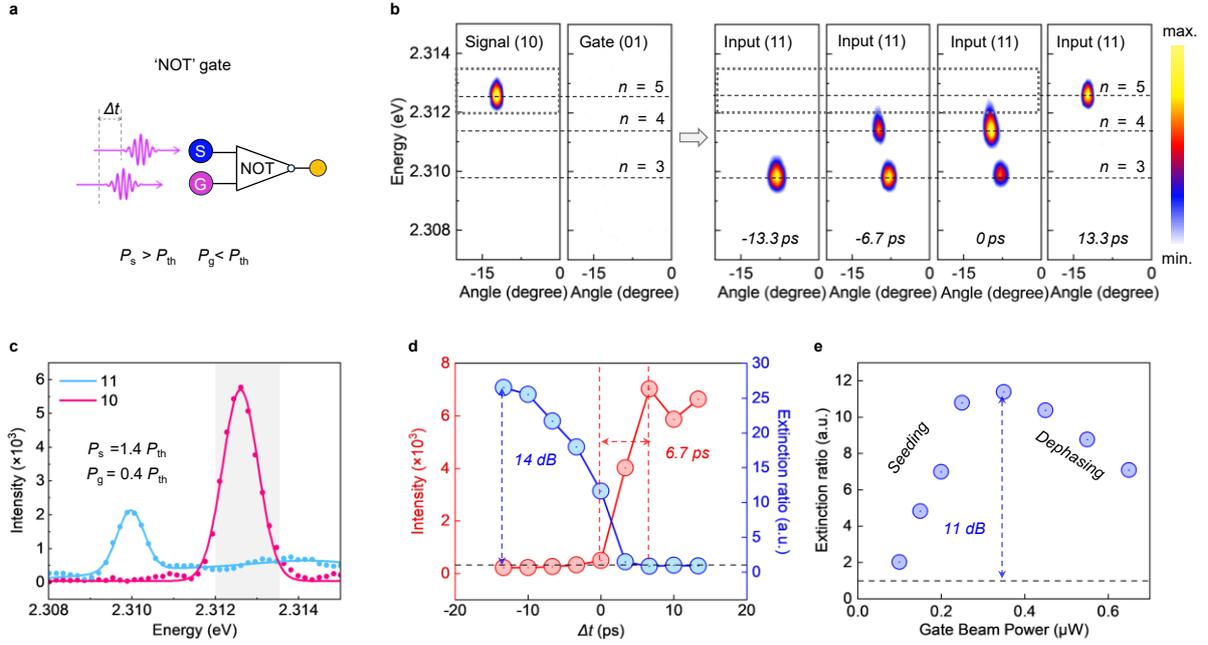

**Figure 3 Ultrafast exciton-polariton 'NOT' logic operation.** (a) Schematic illustration of the 'NOT' gate mechanism, with only the signal beam set above the condensation threshold. (b) ARPL spectra at different time delays, revealing the seeding-induced redistribution underlying the 'NOT' logic operation. The left panels correspond to excitation by the signal beam only ($P_s$ = 1.4 $P_{th}$) and the gate beam only ($P_g$ = 0.4 $P_{th}$), respectively, while the right panels show the condensate distributions at time delays of -13.3 ps, -6.7 ps, 0 ps, and 13.3 ps. The black dashed box marks the collection region corresponding to the 'NOT' gate. (c) Output of the 'NOT' logic gate under different input states. The gray shaded region corresponds to the energy collection window in (b). (d) Emission intensity and extinction ratio as functions of time delay, with the maximum on/off ratio of 14 dB achieved at negative delay times. The intensity shows a rapid decay within 6.7 ps. (e) Extinction ratio as a function of gate beam power at $\Delta t$ = 0 ps, with a maximum extinction ratio of 11 dB.

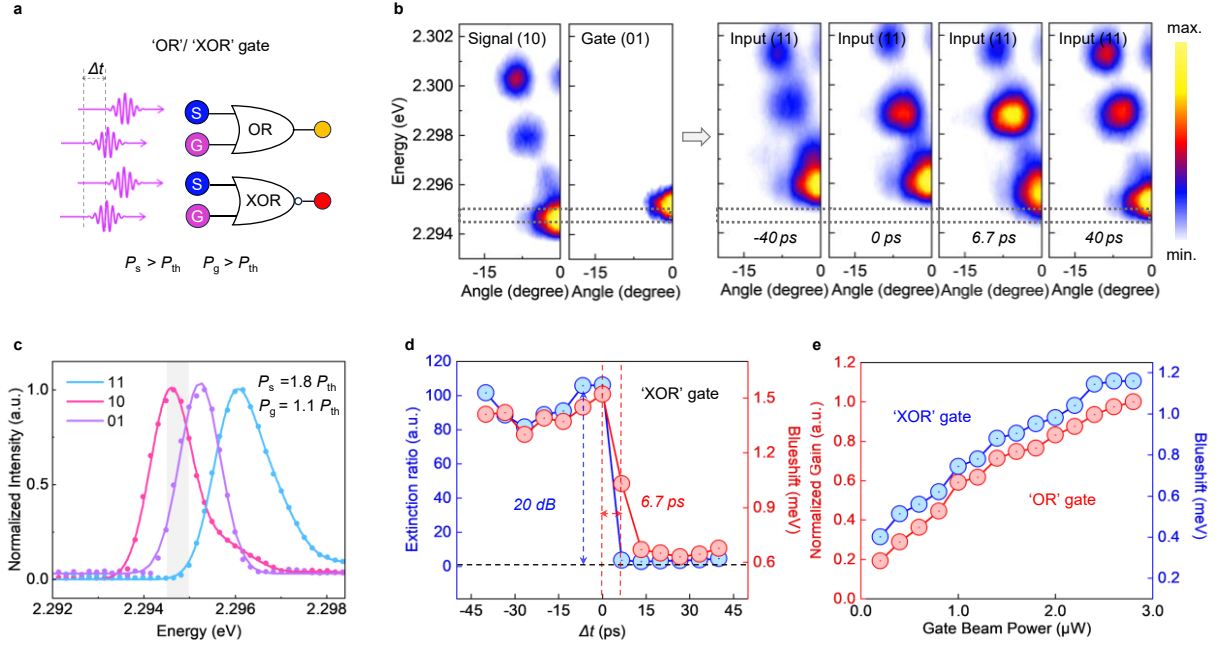

**Figure 4 Ultrafast exciton-polariton 'OR' and 'XOR' logic operations. (a)** Schematic illustration of the 'OR' and 'XOR' gate mechanisms. Both the signal and gate beams are set above the condensation threshold. **(b)** ARPL spectra at different time delays, revealing the nonlinear blueshift underlying the 'OR' and 'XOR' logic operations. The left panels correspond to excitation by the signal beam only ($P_s = 1.8\ P_{th}$) and the gate beam only ($P_g = 1.1\ P_{th}$), respectively, while the right panels show the condensate distributions at time delays of -40 ps, 0 ps, 6.7 ps, and 40 ps. The black dashed box marks the collection region corresponding to the 'XOR' gate, while the entire energy range corresponds to the 'OR' gate. **(c)** Output of the 'OR' and 'XOR' logic gates under different input states. The gray shaded region corresponds to the energy collection window of 'XOR' gate in (b). **(d)** Time-resolved analysis of the condensate characteristics extracted from Gaussian fits. The energy blueshift and extinction ratio are plotted as functions of time delay, showing a maximum extinction ratio of ~20 dB and an ultrafast switching time within 6.7 ps. **(e)** Energy blueshift of the $n = 1$ state (blue line, 'XOR' gate) and normalized gain as functions of gate beam power, where the gain is defined as the ratio of the integrated emission intensity over the entire energy range with and without the gate beam (red line, 'OR' gate).

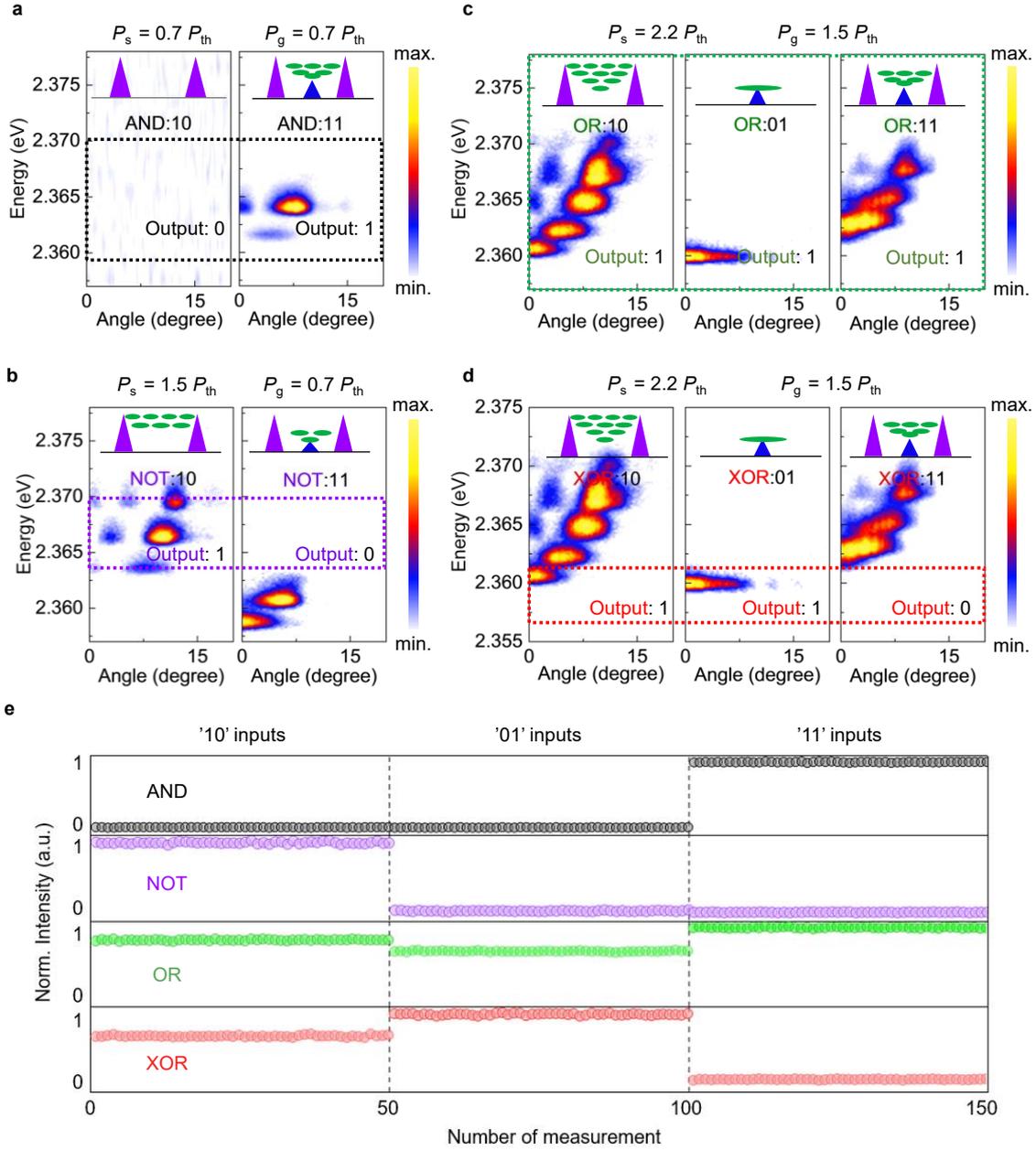

**Figure 5 Multifunctional and reconfigurable exciton-polariton logic gates in a single one-dimensional perovskite microwire.** Insets show conceptual exciton potentials induced by the ring-shaped signal beam (purple triangles) and the laser spot (blue triangles), where the relative height of each triangle indicates the corresponding polariton potential. Green dots represent different polariton states. **(a)** ARPL spectra measured under excitation of $P_s = 0.7\ P_{th}$ and $P_g = 0.7\ P_{th}$. The black dashed box marks the collection region corresponding to the 'AND' gate. **(b)** ARPL spectra measured under excitation of $P_s = 1.5\ P_{th}$ and $P_g = 0.7\ P_{th}$. The purple dashed box marks the collection region corresponding to the 'NOT' gate. **(c, d)** ARPL spectra obtained under the same excitation configuration ($P_s = 2.2\ P_{th}$, $P_g = 1.5\ P_{th}$), distinct detection windows enable the realization of two different logic functions. The green and red dashed boxes mark the regions corresponding to the 'OR' and 'XOR' gates, respectively. **(e)** Normalized output intensities for the 'AND', 'NOT', 'OR' and 'XOR' gates, extracted from the dashed regions in (a)-(d), measured for 50 times.

**Reference**

1. Hopfield, J. J. Theory of the contribution of excitons to the complex dielectric constant of crystals. *Phys. Rev.* **112**, 1555–1567 (1958).
2. Weisbuch, C., Nishioka, M., Ishikawa, A. & Arakawa, Y. Observation of the coupled exciton-photon mode splitting in a semiconductor quantum microcavity. *Phys. Rev. Lett.* **69**, 3314–3317 (1992).
3. Kasprzak, J. *et al.* Bose–Einstein condensation of exciton polaritons. *Nature* **443**, 409–414 (2006).
4. Byrnes, T., Kim, N. Y. & Yamamoto, Y. Exciton–polariton condensates. *Nat. Phys.* **10**, 803–813 (2014).
5. Trypogeorgos, D. *et al.* Emerging supersolidity in photonic-crystal polariton condensates. *Nature* **639**, 337–341 (2025).
6. Christopoulos, S. *et al.* Room-temperature polariton lasing in semiconductor microcavities. *Phys. Rev. Lett.* **98**, 126405 (2007).
7. Kéna-Cohen, S. & Forrest, S. R. Room-temperature polariton lasing in an organic single-crystal microcavity. *Nat. Photonics* **4**, 371–375 (2010).
8. Li, F. *et al.* From excitonic to photonic polariton condensate in a ZnO-based microcavity. *Phys. Rev. Lett.* **110**, 196406 (2013).
9. Tang, J. *et al.* Room temperature exciton–polariton Bose–Einstein condensation in organic single-crystal microribbon cavities. *Nat. Commun.* **12**, 3265 (2021).
10. Wei, M. *et al.* Optically trapped room temperature polariton condensate in an organic semiconductor. *Nat. Commun.* **13**, 7191 (2022).
11. Jin, F. *et al.* Exciton polariton condensation in a perovskite moiré flat band at room temperature. *Sci. Adv.* **11**, eadx2361 (2025).
12. Amo, A. *et al.* Exciton–polariton spin switches. *Nat. Photonics* **4**, 361–366 (2010).
13. Tosi, G. *et al.* Sculpting oscillators with light within a nonlinear quantum fluid. *Nat. Phys.* **8**, 190–194 (2012).
14. Chen, F. *et al.* Optically controlled femtosecond polariton switch at room temperature. *Phys. Rev. Lett.* **129**, 057402 (2022).
15. Genco, A. *et al.* Femtosecond switching of strong light-matter interactions in microcavities with two-dimensional semiconductors. *Nat. Commun.* **16**, 6490 (2025).
16. Ballarini, D. *et al.* All-optical polariton transistor. *Nat. Commun.* **4**, 1778 (2013).
17. Antón, C. *et al.* Quantum reflections and shunting of polariton condensate wave trains: Implementation of a logic AND gate. *Phys. Rev. B* **88**, 245307 (2013).
18. Zasedatelev, A. V. *et al.* A room-temperature organic polariton transistor. *Nat. Photonics* **13**, 378–383 (2019).
19. Piccione, B., Cho, C.-H., Van Vugt, L. K. & Agarwal, R. All-optical active switching in individual semiconductor nanowires. *Nat. Nanotechnol.* **7**, 640–645 (2012).
20. Mirek, R. *et al.* Neuromorphic binarized polariton networks. *Nano Lett.* **21**, 3715–3720 (2021).
21. Li, H. *et al.* All-optical temporal logic gates in localized exciton polaritons. *Nat. Photonics* **18**, 864–869 (2024).
22. Sannikov, D. A. *et al.* Room temperature, cascadable, all-optical polariton universal gates. *Nat. Commun.* **15**, 5362 (2024).

23. Misko, M. *et al.* Temporal bandwidth of consecutive polariton condensation. *Phys. Rev. B* **111**, L161403 (2025).
24. Gao, T. *et al.* Polariton condensate transistor switch. *Phys. Rev. B* **85**, 235102 (2012).
25. Lee, S. W., Lee, J. S., Choi, W. H., Choi, D. & Gong, S.-H. Ultra-compact exciton polariton modulator based on van der waals semiconductors. *Nat. Commun.* **15**, 2331 (2024).
26. Zhao, J. *et al.* Room temperature polariton spin switches based on van der waals superlattices. *Nat. Commun.* **15**, 7601 (2024).
27. Shan, H. *et al.* Tuning relaxation and nonlinear upconversion of valley-exciton-polaritons in a monolayer semiconductor. *Nat. Commun.* **16**, 9700 (2025).
28. Tassan, P. *et al.* Integrated, ultrafast all-optical polariton transistors with sub-wavelength grating microcavities. *Light Sci. Appl.* **15**, 65 (2026).
29. Li, Y. *et al.* Manipulating polariton condensates by rashba-dresselhaus coupling at room temperature. *Nat. Commun.* **13**, 3785 (2022).
30. Liang, J. *et al.* Polariton spin hall effect in a rashba–dresselhaus regime at room temperature. *Nat. Photonics* **18**, 357–362 (2024).
31. Du, W., Zhang, S., Zhang, Q. & Liu, X. Recent progress of strong exciton–photon coupling in lead halide perovskites. *Adv. Mater.* **31**, 1804894 (2019).
32. Song, J. *et al.* Room-temperature continuous-wave pumped exciton polariton condensation in a perovskite microcavity. *Sci. Adv.* **11**, eadr1652 (2025).
33. Peng, K. *et al.* Room-temperature polariton quantum fluids in halide perovskites. *Nat. Commun.* **13**, 7388 (2022).
34. Bao, W. *et al.* Observation of Rydberg exciton polaritons and their condensate in a perovskite cavity. *Proc. Natl. Acad. Sci. U.S.A.* **116**, 20274–20279 (2019).
35. Shang, Q. *et al.* Role of the exciton–polariton in a continuous-wave optically pumped $CsPbBr_3$ perovskite laser. *Nano Lett.* **20**, 6636–6643 (2020).
36. Su, R. *et al.* Perovskite semiconductors for room-temperature exciton-polaritonics. *Nat. Mater.* **20**, 1315–1324 (2021).
37. Wu, X. *et al.* Exciton polariton condensation from bound states in the continuum at room temperature. *Nat. Commun.* **15**, 3345 (2024).
38. Kędziora, M. *et al.* Predesigned perovskite crystal waveguides for room-temperature exciton–polariton condensation and edge lasing. *Nat. Mater.* **23**, 1515–1522 (2024).
39. Feng, J. *et al.* All-optical switching based on interacting exciton polaritons in self-assembled perovskite microwires. *Sci. Adv.* **7**, eabj6627 (2021).
40. Zhu, C. *et al.* Multifunctional integration of laser, waveguide, and all-optical logic gate within a single ultralong $CsPbBr_3$ nanowire. *Nano Lett.* **26**, 780–786 (2026).
41. Wang, J. *et al.* Spontaneously coherent orbital coupling of counterrotating exciton polaritons in annular perovskite microcavities. *Light. Sci. Appl.* **10**, 45 (2021).
42. Tao, R. *et al.* Halide perovskites enable polaritonic XY spin hamiltonian at room temperature. *Nat. Mater.* **21**, 761–766 (2022).
43. Zhang, Y. *et al.* All-optical and ultrafast control of high-order exciton-polariton orbital modes. *Nano Lett.* **25**, 8352–8359 (2025).
44. Kavokin, A. *et al.* Polariton condensates for classical and quantum computing. *Nat. Rev. Phys.* **4**, 435–451 (2022).